# On a Method of Treating Polar-Optical Phonons in Real Space

D. K. Ferry

School of Electrical, Computer, and Energy Engineering
Arizona State University, Tempe, AZ 85287-5706 USA

E-mail: ferry@asu.edu

**Abstract**
Polar-optical phonon interactions with carriers in semiconductors are long range interactions due to their Coulombic nature. Generally, if one wants to treat these with non-equilibrium Green's functions, this long-range interaction requires two- and three-particle Green's functions to be evaluated by e.g. the Bethe-Salpeter equation. On the other hand, optical phonon scattering is thought to be phase-breaking, which, if true, would eliminate this concern over long-range interactions. In seeking to determine just to what extent phase breaking is important, one could treat the polar modes as a real space potential, as is done for impurities, and examine the occurrence of any such correlations. This latter approach suffers from the condition that it is not really known how to handle the polar modes in real space--no one seems to have done it. Here, such an approach is described as one possible method.

## 1. Introduction

The role of the polar-optical phonon in transport in condensed matter systems is relatively old, having been introduced already in early studies of dielectric breakdown [1,2]. Subsequently, it was introduced as a significant scattering process in semiconductor transport [3]. In fact, it is now known that it may well be the dominant scattering process in compound semiconductors such as the III-V materials [4]. Indeed, this scattering process is typically treated through the normal first-order time-dependent perturbation theory that yields the Fermi golden rule for scattering. And, this has been used in nearly all classical and semi-classical treatments of transport whenever the Boltzmann equation is utilized [5].

In recent years, however, quantum effects have become important in semiconductors and in semiconductor devices. As a result, the study of transport in these systems has led to the use of the non-equilibrium Green's functions (NEGF) [6,7], which tend to be extremely difficult computationally. In contrast to most scattering processes, which are localized in space, polar-optical mode scattering is Coulombic in nature [8], as it arises from the polarization between the two atoms of the basic unit cell in a zinc-blende material such as GaAs [9]. The Coulombic scattering in the polar-optical phonon mode is a long-range interaction, and can lead to correlations and interference effects. This already has been shown for impurity scattering [10,11]. Such correlations imply that there are interference effects which do not occur in the semi-classical world, as the phase of the wave functions remains important in such scattering. Naturally, difficulties arise when these scatterers are introduced perturbatively, as they require high-order diagrams appropriate for two- and three-particle Green's function. This, in turn, often leads to the necessity of an evaluation of the Bethe-Salpeter equation for determination of mobility and transport [12,13]. In addition, disconnected diagrams, which have normally been neglected are known to produce variations in the phase of the wave function (in contrast to equilibrium Green's functions, the disconnected graphs do not normally cancel). As a result, the difficulty level in the NEGF computation is raised significantly amount [14].

On the other hand, it has generally been believed that optical-phonon scattering of any type is actually "phase breaking" [15]. This means that long-range correlations that would be expected to exist in the phase of the wave function are destroyed. If this is actually true, there may be no need to proceed to the inclusion of the Bethe-Salpeter equation and multiple-particle Green's functions. That is, if the interactions do break the phase, then one may need only consider lowest order graphs.

In the analogous situation of impurity scattering, which is also a Coulombic interaction, the long-range correlations were discovered when treating the interaction as a real-space scattering process; e.g., the carrier was deflected by the real-space Coulomb potential, rather than by a perturbative approach. This has also been shown to work well for the carrier-carrier many-body interactions [16]. Consequently, the existence of long-range correlations in a process that is supposedly phase breaking may be examined more closely in such real-space processes. The obvious limitation to such an approach lies in the fact that the phonon modes have never been considered in this manner. Phonons are a cooperative motion of the atoms in the atomic lattice, and are not thought to be localized at a point in space. But, it is not clear that this can actually be done, although the recent treatment of even phonons as deterministic particles [17,18,19] suggests that it is possible to treat them as localized objects.

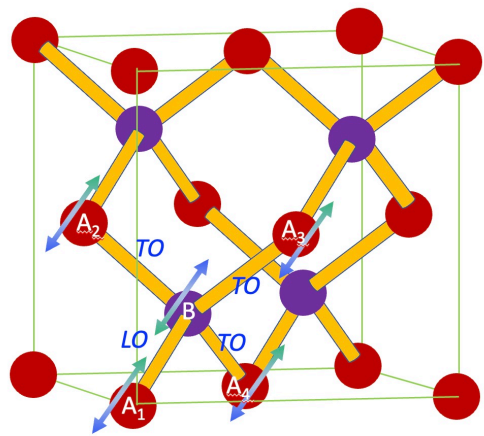

**Fig. 1** The face-centered unit cell of the zinc-blende lattice. The two basis is indicated by the atoms labeled $A_1$ and B. The LO phonon motion is indicated by the blue arrows.

In this paper, a possible approach to treating the polar-mode phonons in real space by their polarizations localized in space. The polar modes have very small group velocities, so these scattering centers are treated as localized oscillating potentials rather than as extended modes or even as particles. In Sec. 2, the nature of the polarizations and potentials is described to clarify how these localized quantities will be determined. Section 3 deals with the nature of a Monte Carlo approach and the problem of energy transfer. Finally, in Sec. 4, some conclusion and observations will be presented.

## 2. Localized Polar-Optical Phonons

In Fig. 1, the motion of the two atoms on the basis of the face-centered cubic lattice is shown by the blue arrows around the two basis atoms labeled as $A_1$ and B. This motion corresponds to the longitudinal polar-optical phonon. Note that atom B is tetrahedrally coordinated also to atoms $A_2$-$A_4$, but motion between these atoms and the B atom is actually the transverse optical (TO) phonon. This latter connection keeps the various LO modes on the A atoms coordinated and leads to the long-range coherence of the phonon wave. The phonon dispersion for GaAs, a typical III-V semiconductor with the zinc-blende lattice, is shown in Fig. 2 for convenience. The LO (polar) and TO modes are indicated and plotted in green and red, respectively. The acoustic, which are not of interest here, are shown in black.

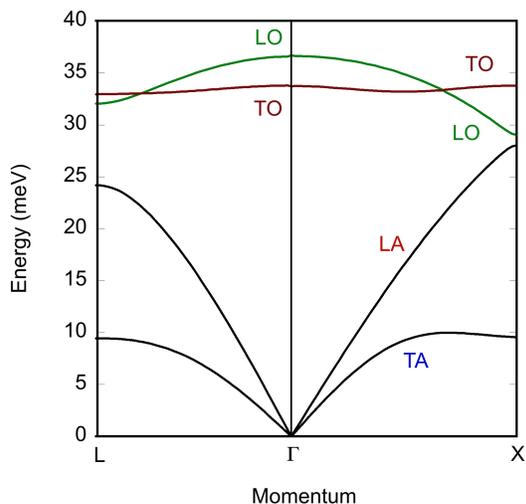

**Fig. 2** The phonon spectrum for GaAs.

The splitting between the LO and TO modes at the $\Gamma$ point arises from the ionicity of the lattice and the charge dipole that exists between the two atoms of the basis of the lattice. The polar-optical (LO) phonon differs in energy from the TO mode due to an electromagnetic response. In III-V compounds, the atoms in the face-centered cubic lattice are tetrahedrally coordinated, just as in the group IV materials. Each atom has four nearest neighbor atoms to which it is covalently bonded. Hence, on average each atom has four bonding electrons. In the III-V compounds, however, one atom has only three electrons and the other has five. While the average atom still has four bonding electrons, this requires charge transfer from the group V to the group III atom.

The amount of average charge on each of the two atoms is known as the *effective charge,* although there have been several forms of effective charge suggested [20, 21]. The resulting dipole between the two neighboring atoms interacts with optical waves in the far infrared, and creates a difference between the optical dielectric function in the visible ($\epsilon_\infty$) and the static, or low frequency, dielectric function ($\epsilon_s$). This interaction is characterized as a polarization related to the exrternal applied flux density **D** as [22]

$$\boldsymbol{P} = \left(\frac{1}{\epsilon_\infty} - \frac{1}{\epsilon_s}\right)\boldsymbol{D}, \qquad (1)$$

where **D** is the electric flux density. That is, the dipole is characterized by the two dielectric functions that actually convert the charge $e$ to the effective charge $e^*$. As may be seen in Fig. 1, the dipole is oriented along a (111) direction, and there are four such directions (the reverse directions yield the same effect) in the crystal.

There are two views as to how to treat the individual dipoles in real space. In one view, the Bose-Einstein distribution

$$f_{BE} = \left[e^{\hbar\omega_0/k_B T} - 1\right]^{-1} \qquad (2)$$

gives the number of phonons of energy $\hbar\omega_0$, where $\omega_0$ is the radian frequency of the LO mode at $\Gamma$, in this particular Fourier component (the number of excitations of the harmonic oscillator representation of this mode). This view transfers the occupancy given by (2) to the amplitude of this particular phonon mode.

An alternative view, which connects better with the real-space approach, is to consider that (2) gives the actual probability that a particular dipole is actually excited. That is, the product of (2) with the 3D density of states for phonons of energy $\hbar\omega_0$ gives the number of dipoles per unit volume that are excited. Thus, for example, in GaAs the phonon density of states is combined with the linewidth at the LO phonon frequency to give the number of dipoles as [23]

$$n_d = D(\omega)\Delta\omega \sim \frac{q^2}{2\pi^2}\frac{\Delta\omega}{\partial\omega/\partial q} \ . \qquad (3)$$

The momentum vector involved in elecctron scattering is approximately $\sqrt{2m^*\omega_0/\hbar}$, or about $2.4 \times 10^8$ /m in GaAs, and the group velocity may be determined from the phonon spectrum, giving about 130 *m/s* for the LO mode. The linewidth at LO mode near $\Gamma$ is about $8.3 \times 10^{10}\ s^{-1}$. This gives an estimate of the density of states for this mode to be something like $1.7 \times 10^{24}\ m^{-3}$, and (2) gives a probability of 0.32, so that the number of dipoles excited at 300 K would be something like $5 \times 10^{23}\ m^{-3}$. This implies that each dipole occupies a volume of some 2000 *nm*³. Thus, it is relatively easy to construct a structure through which a quantum wave can propagate and the nature of the resulting quantum effects evaluated. Using the equivalent form for two dimensions, by assuming quantum well of ~10 nm thickness, one arrives at an area of ~200 *nm,*² which is slightly smaller than the area per impurity used in the Wigner study of quantum interference with impurities [11]. Thus, it seems quite reasonable that one could do a similar study of real-space scattering from a polar-optical mode dipole.

### 3. Form of the Dipole

Dipoles are a common textbook problem in electricity and magnetism. When the dipole is aligned along the polar axis, then the potential may be written as

$$V(r) \sim \frac{d}{4\pi\epsilon r}cos\vartheta \ , \qquad (4)$$

where d is the dipole ($= 2e^*a$) and *a* is the separation of the two atoms in the basis. This form assumes that the view is relatively far from the atomic level. But, the phonon is not the static atomic separation, as this *static* dipole is already included in the band structure calculation. Rather, it is the oscillatory motion of the atoms around their static positions that yield the phonons. In this case, however, (4) is an oscillating potential, so that the real part may be expressed as

$$V(r) \sim \frac{d}{4\pi\epsilon r}cos\vartheta cos(\omega_0 t) \ . \qquad (5)$$

That is, the potential is reversing itself in sign at the polar-optical mode frequency, which is of the order of 9 THz in GaAs. While the static dipole may deflect an electron, one must remember that it is part of the ensemble of atoms that make the crystal structure, and its role (through the band structure) contributes to the effective mass. Scattering arises from the time varying a.c. part of the dipole. Certainly, the amplitude of the atomic motion is small compared to a normal dimension involved in carrier transport.

It is important to ask whether or not a passing electron will actually see the oscillating potential. This is easy to estimate. It was noted above that, in a two-dimensional case, each dipole occupied about $10^3$ *nm*², which may be taken to be a square ~32 nm on a side. An electron traveling at ~ $3 \times 10^7 cm/s$ (approximately the thermal velocity at room temperature) would traverse this distance in about 0.1 ps, which is roughly the period of the oscillation. Thus, it would certainly have adequate time to interact with the potential. It would thus be deflected in a manner similar to any other Coulomb potential, although this would be moderated somewhat by the oscillatory nature of (5).

The amplitude of the potential, the effective dipole term in (4), can be estimated from considerations of the energy in the corresponding harmonic oscillator. The energy in the oscillator can be taken to be [24]

$$\begin{aligned} E &= \left(n + \frac{1}{2}\right)\hbar\omega_0 \\ &= \frac{P^2}{2M} + \frac{1}{2}M\omega_0^2 x^2 \end{aligned} \ , \qquad (6)$$

where $P$ is the atomic momentum, $x$ is the atomic displacement and $M$ is the reduced mass of the two atoms. As is common for harmonic oscillators, whether classical or quantum mechanical, one-half of the energy is in each of the two terms in the second line of (6). Hence, one can estimate the average oscillatory displacement of the atoms as

$$x = \sqrt{\left(n + \frac{1}{2}\right)\frac{\hbar}{M\omega_0}} \ . \quad (7)$$

For GaAs, using values described above and a relative reduced mass of ~36 gives a displacement of approximately 5 pm. This gives a magnitude for the double atom system of 10 pm, or about 4% of the nearest neighbor distance. While small, this is certainly doable with reasonable precision is a real space simulation.

## 4. The Problem of Energy Exchange

To this point, the approach is relatively easy. However, in scattering by the polar-optical phonon dipole, there is an energy exchange, corresponding to emission or absorption of a phonon. This is the first problem that has not already been addressed with the treatment of impurity and carrier-carrier scattering in real space. In GaAs, this energy is about 36 meV, and a decision about how to handle this energy exchange needs to be made.

It has been known for some time that it requires a few fs in order for the carrier to actually scatter and emit, or absorb, the phonon energy [25,26]. If one assumes a time of some 2 fs, and the above mentioned velocity for the carrier, this carrier will drift almost a nm during the actual collision. This is already considerably larger than the motion of the dipole. Thus, theoretically, the carrier is losing, or gaining, the energy during this entire transit past the dipole. Statistically, over the occurence of several scattering events, one could enforce the energy exchange anywhere during this transit, but the easiest approach would be to have it occur at the point of closest approach to the dipole.

There is a further point to consider. Since the probability of there being a phonon in the dipole was stated earlier as 0.32, it is unlikely that the carrier can actually absorb a phonon, when there is inadequate energy in the dipole. This would suggest that absorption cannot occur in this situation (temperature, etc.) until one has been emitted by the carriers. Only then can emission and absorption begin to establish detailed balance in an equilibrium situation. This would mean that the dipole has to be driven out of equilibrium by this first emission process. The emission of the phonon energy would raise the occupancy probability (2) from 0.32 to 1.32. This would, in turn, increase the amplitude of the oscillation from about 10 pm to about 13.6 pm, or a 36% increase.

The lifetime of a non-equilibrium phonon in GaAs is about 3 ps [27]. With the group velocity determined in section 2, this excess phonon can drift only about 0.4 nm, and could diffuse a much smaller distance. This is slightly larger than the nearest-neighbor distance between the atoms, but suggests that this excess phonon isn't going anywhere. It is going to remain localized at the dipole with which the carrier interacted. Hence, the non-equilibrium phonon population throughout the semiconductor is localized on the dipole sites, and relatively independent of wave momentum, at least up to the order of $10^9 \ m^{-1}$. But, the phonon momentum discussed in section 2 for the polar-optical mode near Γ is not that much smaller than this value.

## 4. Discussion

Treating a scattering process in real space is a method of discovering quantum effects lead to interference around the scattering. While such interference is known classically for optical waves, it is not normally seen for particles. That is, until the particles are treated as waves in their quantum existence. It is not generally appreciated that particles can lead to interference just as optical waves. That is, the typical two slit experiment may be reproduced using particles, and has been for electrons in a transmission electron microscope (TEM). The use of a biprism, a charged wire in the TEM acts to create the two slits (on either side of thewire) and generates a typical interference pattern [28]. This is also observed to occur in semiconductors for a single charged impurity [11], and two impurities lead to a three-slit interference pattern [11]. Thus, the purpose of treating the polar-optical phonons in real space is to determine whether or not such interference patterns form in the scattered wave patterns. If they do form, then the concept of phase breaking is inaccurate.

Phase breaking is a process in which the wave function's phase is randomized by the scattering interactions. This is thought to occur due to the energy change (and hence the frequency) within the phase of a quantum mechanical object. Since the interference occurs in real space, it is natural to then seek to examine the scattering process in real space as done with impurity scattering [10,11]. To my knowledge, this has never been done with optical phonon scattering. Since, such interference is more likely with Coulombic (long range) interactions, the polar-optical phonons are the most likely optical mode in which to seek this interaction.

It has been shown here that the parameters and range of the interaction is within a set of doable parameters with which to treat the interaction in real space. As a result, a possible method of approach to doing so has been described in this paper. Such a similation will go a long way toward further understanding of the phase breaking process, and answer critical questions about is role in quantum physics.